# Observation of the magnetic spin Hall effect in a ferromagnet

Nicholas Davey-García[1,2,†], Jone Mencos[1,2,†], Luciano Bravo[1,2], Inge Groen[3], Luis E. Hueso[1,4], Andreas Berger[1], Fèlix Casanova[1,4,*]

[1] CIC nanoGUNE BRTA, 20018 Donostia-San Sebastián, Basque Country, Spain
[2] Departamento de Polímeros y Materiales Avanzados: Física, Química y Tecnología, University of the Basque Country (UPV/EHU), 20018 Donostia-San Sebastián, Basque Country, Spain
[3] Univ. Grenoble Alpes, CEA, CNRS, Grenoble-INP, SPINTEC, 38000 Grenoble, France
[4] IKERBASQUE, Basque Foundation for Science, 48009 Bilbao, Basque Country, Spain
[†]Equal contribution
[*]E-mail : f.casanova@nanogune.eu

### Abstract

The conventional spin Hall effect generates spin currents whose flow direction, spin polarization, and driving electric field are mutually perpendicular. Magnetic order lifts this symmetry restriction and enables additional time-reversal-symmetry-odd components of the spin-conductivity tensor, giving rise to the magnetic spin Hall effect (MSHE). These components also couple the generated spin polarization to the magnetic order, providing a degree of control absent in the conventional spin Hall effect. Although the MSHE has been observed in antiferromagnets, its experimental identification in conventional ferromagnets has remained elusive. Here, using a non-local lateral spin-valve geometry, we electrically identify the MSHE and its reciprocal effect in a perpendicularly magnetized Co-based multilayer. Reversal of the multilayer magnetization reverses the MSHE and magnetic inverse spin Hall signals, revealing their time-reversal-symmetry-odd character and magnetization control. By contrast, the conventional spin Hall and inverse spin Hall signals measured in the same devices remain unchanged under magnetization reversal, consistent with their time-reversal-symmetry-even character. We obtain a magnetic spin Hall angle of $\theta_{\mathrm{MSH}} = (3.8 \pm 0.6)\%$, comparable in magnitude to the spin Hall angle of heavy metals commonly used in spintronic devices, such as Pt. These results establish the MSHE as a sizable, magnetically switchable spin-charge interconversion mechanism in conventional ferromagnets.

## INTRODUCTION

Among the various mechanisms available for spin-current generation and detection, the spin Hall effect (SHE) has emerged as a central paradigm since its experimental demonstration two decades ago [1]. In materials with strong spin-orbit coupling (SOC), a longitudinal charge current causes electrons with opposite spin orientations to be deflected in opposite transverse directions, thereby generating a pure spin current. The reciprocal effect, known as the inverse spin Hall effect (ISHE), converts a pure spin current flowing through the material into a transverse charge current [2]. These efficient and scalable means of interconverting charge and spin currents are now ubiquitous in modern spintronic technologies, playing a key role in devices such as magnetic random-access memory (MRAM) [3] and magnetoelectric spin-orbit (MESO) logic [4]. The SHE can be formally described using the third-rank spin conductivity tensor $\sigma_{ij}^{k}$, which relates the spin-current density $J_i^k$ to the electric field $E_j$ through $J_i^k = \sigma_{ij}^k E_j$. Here, $i$ denotes the direction of the spin-current flow, $j$ the direction of the applied electric field, and $k$ the spin-polarization axis of the transported spins. In nonmagnetic metals with cubic symmetry, the tensorial response reduces to a single isotropic quantity, the spin Hall conductivity $\sigma_{SH}$, which is even under time-reversal symmetry (TRS) [5] and requires the electric field, spin-current flow and spin polarization to be mutually perpendicular, as illustrated in Fig. 1a.

More recently, several studies have reported the existence of time-reversal-odd contributions to the spin-conductivity tensor [6,7]. These additional components originate not only from the underlying crystal structure of the material, but also from magnetic ordering, which reduces crystal symmetries and explicitly breaks TRS. Such contributions have been recently discussed in the case of noncollinear [8,9] and collinear [10] antiferromagnets, where they give rise to an effect known as the magnetic spin Hall effect (MSHE) [11,12]. The MSHE, as opposed to the conventional SHE, can generate a transverse spin current when the electric field and the spin polarization are parallel between them, and reverses sign under time reversal, as shown in Fig. 1b. Furthermore, theoretical works predict that the MSHE should also emerge in ferromagnetic metals, such as Fe, Ni and Co [7,13]. While experimental signatures of magnetization-dependent spin-charge interconversion have been reported in yttrium iron garnet/Pt/Co/Pt heterostructures using spin-Seebeck-generated spin currents [14], the unambiguous electrical isolation of the MSHE in ferromagnetic conductors remains elusive. Such isolation requires a measurement geometry in which spurious effects (parasitic magnetoresistive or thermoelectric contributions) do not contaminate the spin signal. This condition is fulfilled in lateral spin-valve (LSV) architectures, where non-local detection enables a direct and unambiguous probing of spin-charge interconversion in a broad class of materials [15–22]. In addition, isolating magnetization-dependent contributions to the SHE demands independent control of the magnetization of the spin detector and of the material hosting the MSHE. In this work, we employ a $(Pt/Co)_n$ multilayer as a prototypical ferromagnet exhibiting strong perpendicular magnetic anisotropy (PMA), which allows for decoupled magnetization switching. Using a $Ni_{80}Fe_{20}/Cu/(Pt/Co)_n$ LSV device, we unequivocally identify the MSHE in a perpendicularly magnetized Pt/Co multilayer by electrically isolating the time-reversal-odd component of its spin-conductivity tensor.

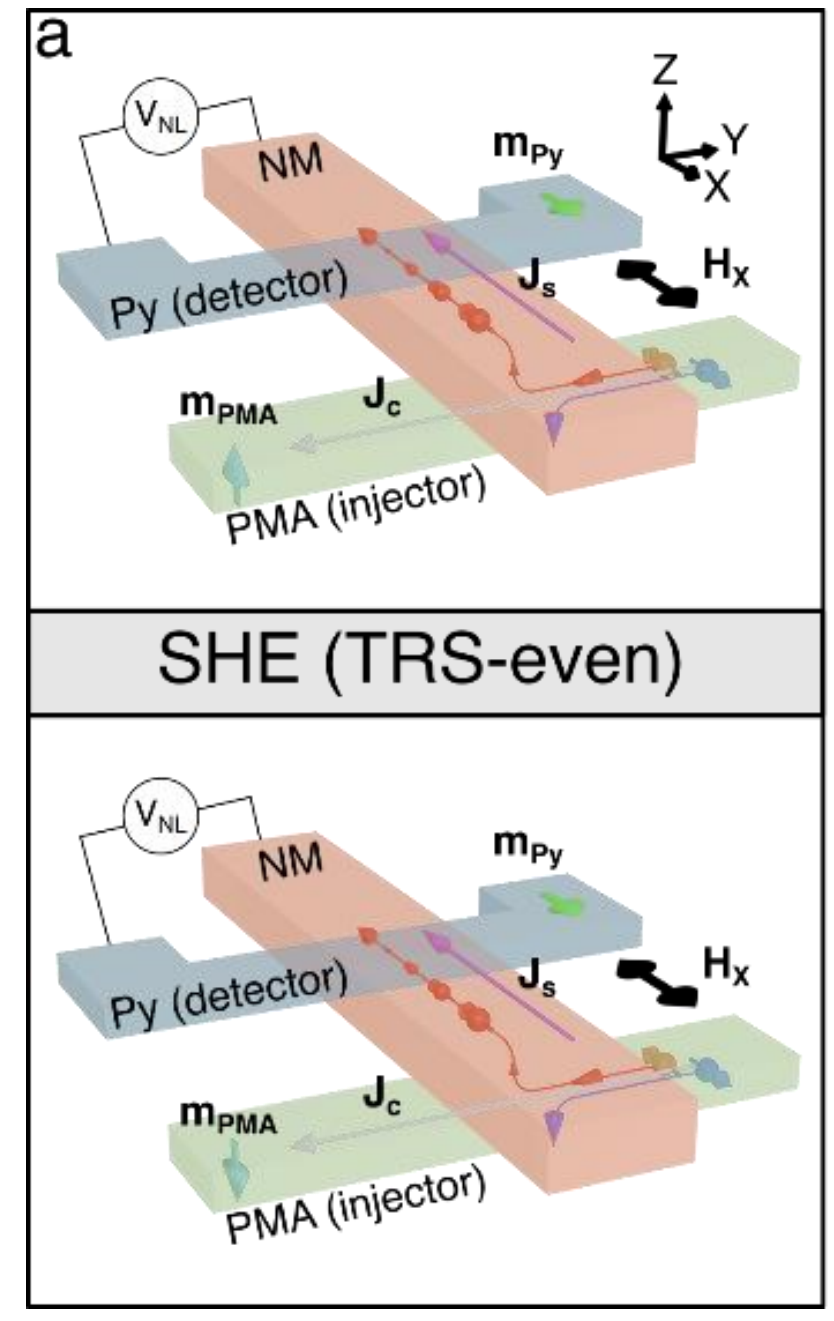

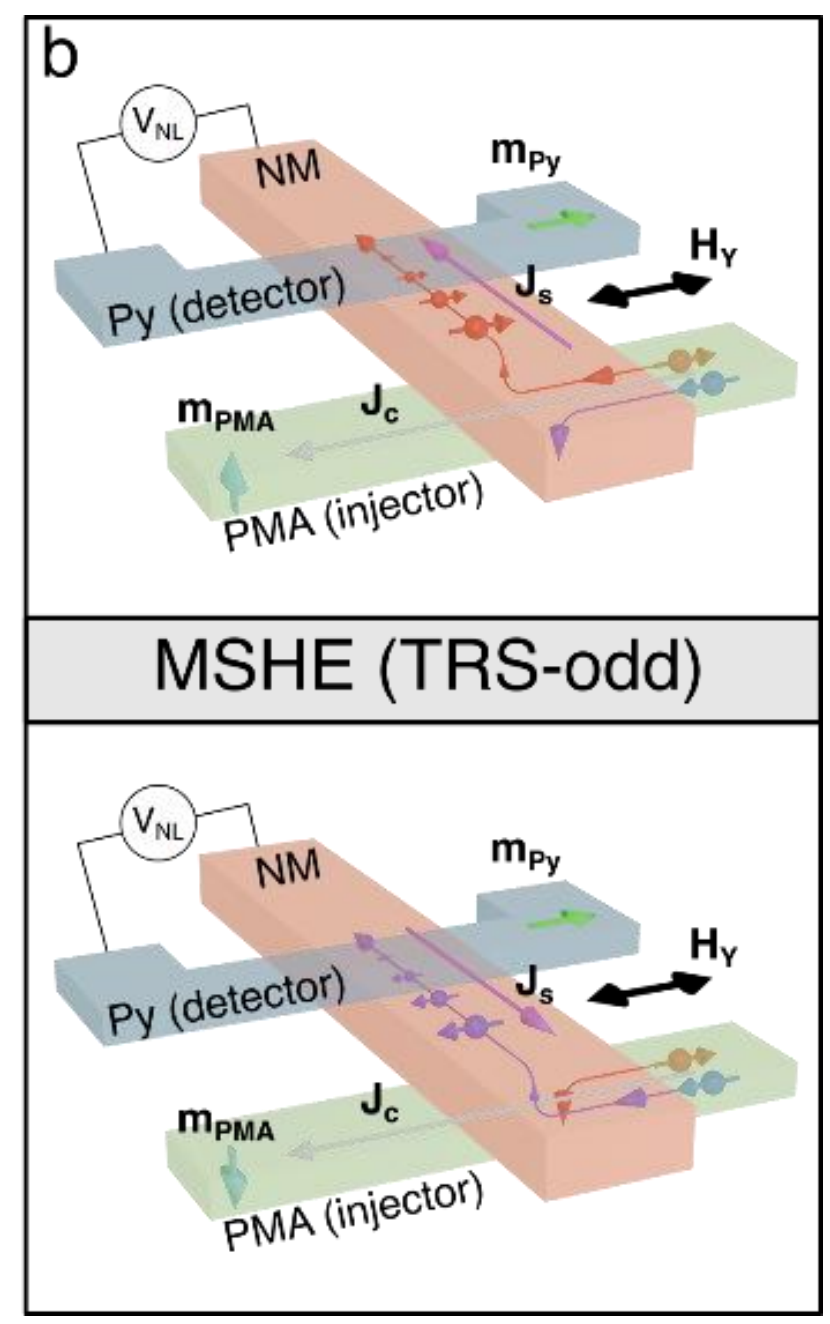

**FIG. 1. Spin Hall effect and magnetic spin Hall effect mechanisms in a lateral spin valve.** (a) Schematics of the LSV showing the SHE in the PMA electrode (green), spin transport in the Cu channel (orange), and detection in the Py electrode (blue). The magnetization $\boldsymbol{m_{PMA}}$ can be oriented along +*z* (top) and -*z* (bottom). The charge current $\boldsymbol{J_c}$, spin current $\boldsymbol{J_s}$ and spin polarization are mutually orthogonal. Reversal of $\boldsymbol{m_{PMA}}$ leaves the sign of the generated spin current unchanged (TRS-even). (b) Schematics of the LSV showing in this case the MSHE in the PMA electrode, for $\boldsymbol{m_{PMA}}$ oriented along +*z* (top) and -*z* (bottom). The spin polarization is parallel to the charge current $\boldsymbol{J_c}$, and both are perpendicular to the spin current $\boldsymbol{J_s}$. Reversal of $\boldsymbol{m_{PMA}}$ reverses the sign of the generated spin current (TRS-odd).

## EXPERIMENTAL DETAILS

$(Co/Pt)_n$ multilayers were grown on $Si/SiO_2$ substrates (n+-doped Si, with a 300-nm-thick thermally grown oxide layer) by DC sputtering with a repetition of n = 10 on a seed layer (Fig. 2a) to achieve a ferromagnetic stack with PMA. A more detailed description of the growth process of the material is included in Supp. Info. Note 1. Subsequently, 100-nm-wide Hall cross electrodes were patterned with negative e-beam lithography and Ar-ion etching to characterize the AHE of the PMA system as well as integrate it into LSVs. Next, 100-nm-wide ferromagnetic $Ni_{80}Fe_{20}$ (permalloy, Py) injection and detection electrodes were patterned on either side of the PMA electrode by positive e-beam lithography, e-beam evaporation of 20 nm of Py and lift-off. 100-nm-wide transverse non-magnetic Cu channels were patterned by positive e-beam lithography, thermal evaporation of 60 nm of Cu and lift-off, connecting the Py and PMA electrodes. The edge-to-edge separation between Py electrodes is 600 nm while the edge-to-edge separation between the PMA and each Py electrode is 250 nm (Fig. 2c). A reference LSV without the PMA electrode is fabricated simultaneously (Fig. 3a). The entire layout of these LSVs allows us to characterize the spin diffusion length of the PMA, as well as to measure and quantify the MSHE and its inverse, the magnetic inverse spin Hall effect (MISHE). This architecture also enables the characterization of the conventional SHE and its inverse, the ISHE. The results for one device are presented and discussed below, but multiple devices have been fabricated and measured, and the extracted parameters have been summarized in Supp. Info. Table I.

Figure 2b presents the out-of-plane room-temperature superconducting quantum interference device (SQUID) magnetometry measurement of the unpatterned stack. The measurement shows a clear hysteresis loop with a coercive field $H_{z,C}\sim$ 0.15 T, confirming the presence of PMA. The measured AHE curve is shown in Fig. 2d when sweeping the magnetic field in the out-of-plane *z*-direction, reproducing the hysteresis of the whole film, with a similar $H_{z,C}$. When the magnetic field is swept along the in-plane *y*-direction starting from the two spontaneous states at zero field, the transverse resistance decreases in a step-like manner toward its baseline value, indicating the occurrence of magnetic domains within the PMA multilayer at intermediate field values prior to in-plane saturation.

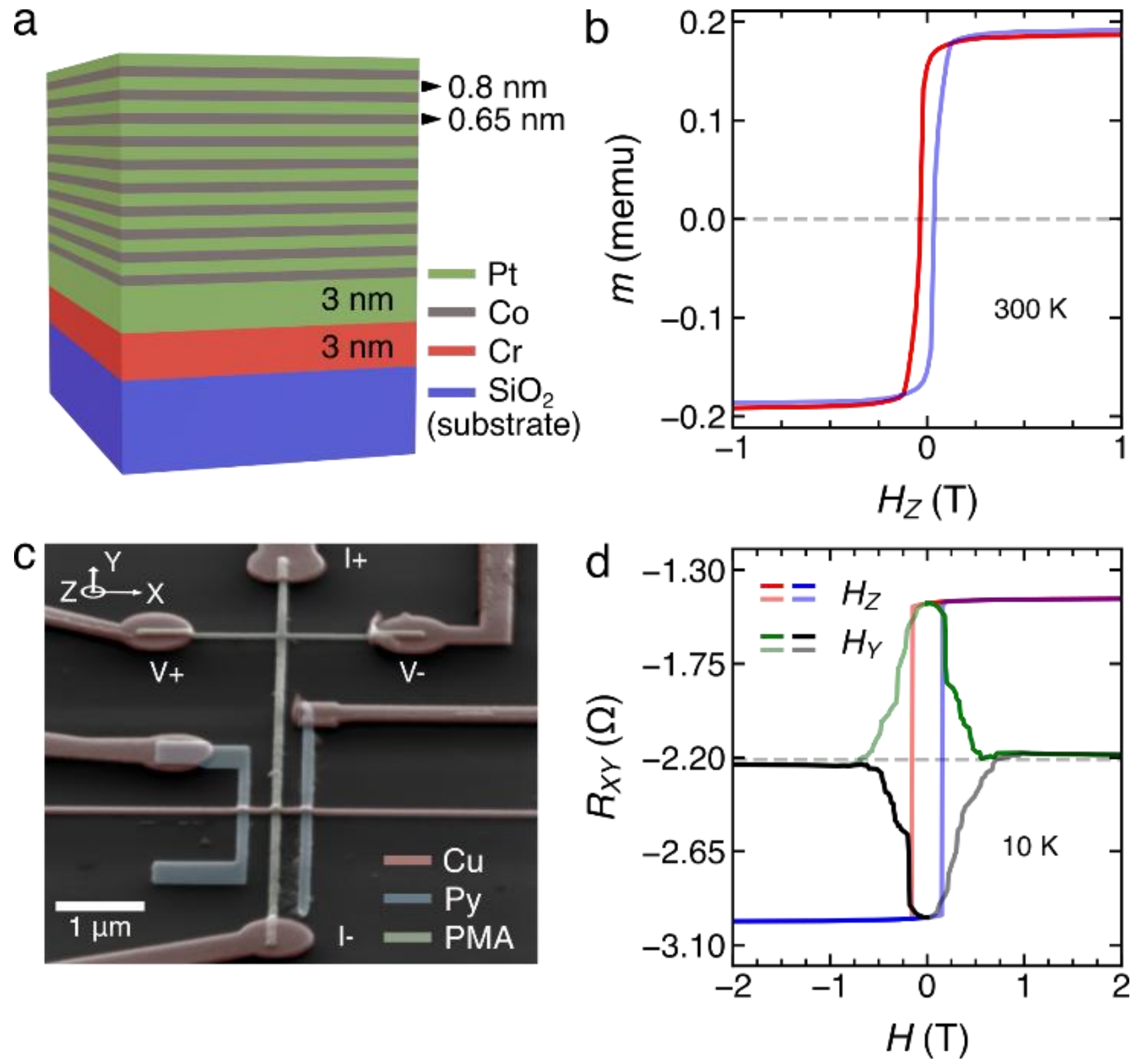

**FIG. 2. Characterization of the ferromagnet with perpendicular magnetic anisotropy.** (a) Schematics of the PMA multilayer structure $(Co/Pt)_n$ with n = 10 grown on a seed layer (3 nm of Cr and 3 nm of Pt). The PMA properties arise from the subsequent stacking of Pt/Co layers. The thicknesses of each layer have been verified by X-ray reflectometry measurements (see Supp. Info. Fig. S1). (b) Out-of-plane magnetization of the PMA ferromagnet at 300 K measured using SQUID magnetometry. (c) False-colored scanning electron microscope image of the PMA multilayer Hall cross electrode, where the electrical configuration for the AHE measurement is labeled. (d) AHE measured in the PMA Hall cross electrode with the configuration shown in panel c. Magnetic field sweeps were performed along the out-of-plane *z*-axis (trace and retrace correspond to red and blue curves, respectively). Magnetic field sweeps along the in-plane *y*-axis were taken from zero field after applying an out-of-plane field of +2 T (dark and light green curves correspond respectively to positive and negative field sweep) and -2 T (grey and black curves correspond respectively to positive and negative field sweep). The grey dashed line indicates the baseline resistance, obtained as the average value between the resistance values with out-of-plane field at saturation.

## RESULTS AND DISCUSSION

In a LSV, when an electrical current ($I_{bias}$) is injected from the Py injector into the Cu channel, a spin accumulation is created at the Py/Cu interface, generating a pure spin current that diffuses along the Cu channel toward the Py detector [23]. The decaying spin current is detected as a non-local voltage ($V_{NL}$) between the Py detector and the Cu channel, with the non-local resistance defined as $R_{NL} = V_{NL}/I_{bias}$. The spin signal $\Delta R_{NL}$, defined as the difference in $R_{NL}$ between the parallel and antiparallel magnetization configurations in the *y*-axis of the two Py electrodes (see black arrows in Fig. 3b), is proportional to the spin current reaching the detector. When an additional electrode is inserted between the injector and detector, it absorbs part of the spin current, thereby reducing $\Delta R_{NL}$. Using the two devices shown in Fig. 3a, we measure the spin signal of the reference LSV (voltage detected between contacts 8 and 1 in the biasing configuration across contacts 9 to 10), $\Delta R_{NL}^{ref}$, and that of the LSV containing the intermediate PMA electrode (voltage detected between contacts 6 and 10 in the biasing configuration across contacts 2 to 1), $\Delta R_{NL}^{abs}$. The corresponding $R_{NL}$ curves are plotted in Fig. 3b. In both devices, the switching of the Py detector and injector is clearly observed at their respective in-plane coercive fields, $H_{y,C}\sim$ 0.03 T and ~0.1 T.

The spin diffusion length of the PMA electrode, $\lambda_{PMA}$, can be obtained from the ratio $\eta = \Delta R_{NL}^{abs}/\Delta R_{NL}^{ref}$. Note that, in a ferromagnetic material, the spin diffusion length exhibits anisotropy related to the relative orientation between the polarization of the spin current and the magnetization of the material [24]. In our measurement, when applying the magnetic field along the easy axis of the Py contact, the relative orientation of the injected spins and the PMA magnetization is perpendicular one to the other, thus probing the value $\lambda_{PMA}^{\perp}$. The relation between the ratio $\eta$ and $\lambda_{PMA}^{\perp}$ is derived using the one-dimensional spin-diffusion model [16,23], assuming transparent interfaces [17] and can be found in Supp. Info. Eq. (S1). All relevant parameters, including the resistivities (Supp. Info. Fig. S2), and spin diffusion lengths of Cu and Py, the spin polarization of Py, and the device dimensions, are determined independently, leaving $\lambda_{PMA}^{\perp}$ as the only free parameter. Measurements from four devices yield a mean spin diffusion length of $\overline{\lambda_{PMA}^{\perp}} = 0.31 \pm 0.06$ nm.

Having characterized the spin absorption by the PMA electrode, we next investigate its ability to generate spin currents via the MSHE, as shown in Fig. 3c. In this case, $I_{bias}$ is applied along the PMA wire (between contacts 4 to 7), generating through the MSHE a transverse spin current flowing out of plane. This spin current is injected into the Cu channel and diffuses towards the Py detector, where it is detected as a non-local voltage ($V_{MSHE}$, between contacts 2 and 1). We define the corresponding MSHE resistance as $R_{MSHE} = V_{MSHE}/I_{bias}$. To satisfy the symmetry conditions for detecting the MSHE, the magnetic field is swept along the in-plane easy axis of the Py electrode (*y*-axis). The magnetization of the PMA electrode is first preset by applying an out-of-plane magnetic field $H_Z = +2$ T and then reducing it back to zero. Subsequently, a magnetic field sweep along y, from −0.05 T to +0.05 T, reverses the magnetization of the Py detector at $H_{y,C}\sim$ 0.03 T, producing the switching of $R_{MSHE}$ shown by the red curve. Throughout this sweep, the magnetization state of the PMA electrode remains unchanged, as established in Fig. 2d. When the PMA magnetization is instead preset in the opposite direction, $H_Z = -2$ T, the detected signal reverses sign, as shown by the blue curve. The observed sign reversal upon reversal of the PMA magnetization demonstrates the TRS-odd character of the spin current generation (Fig. 1b), consistent with an MSHE origin in the PMA electrode.

Measurements of the inverse effect, the MISHE, are shown in Fig. 3d. For these measurements, the voltage and current probes are interchanged relative to the MSHE configuration. In this case, $I_{bias}$ is injected from the Py electrode into the Cu channel, generating a spin current that diffuses toward and is absorbed by the PMA electrode. The absorbed spin current is converted by the MISHE into a transverse charge current, which is detected under open circuit conditions as a voltage ($V_{MISHE}$). The corresponding MISHE resistance is defined as $R_{MISHE} = V_{MISHE}/I_{bias}$. As in the direct MSHE measurement, the MISHE signal also reverses sign when the magnetization of the PMA electrode is reversed, as seen by comparing the red and blue curves in Fig. 3d, confirming its TRS-odd character.

We next compare the direct and inverse configurations. The signal amplitudes $\Delta R_{MSHE}$ and $\Delta R_{MISHE}$, defined as the differences between the corresponding resistances measured for the *+y* and *-y* magnetization states of the Py electrode, are identical in magnitude, $|\Delta R_{MSHE}| = |\Delta R_{MISHE}|$. Moreover, for the same preset magnetization state of the PMA electrode, the direct and inverse signals have the same sign (same-color curves in Figs. 3c and 3d). Both the equality of the amplitudes and the sign relation between the direct and inverse configurations are governed by Onsager-Büttiker reciprocity [25,26] (see Supp. Info. Note 3). Crucially, the full reciprocity operation requires interchanging the current and voltage probes and reversing all time-reversal-odd parameters, including the magnetization of the PMA electrode. Accordingly, the reciprocal pair is obtained by comparing the $R_{MSHE}$ measurement after presetting the PMA magnetization with +2 T (Fig. 3c, red curve) and the $R_{MISHE}$ measurement after presetting it with -2 T (Fig. 3d, blue curve). The opposite signs of these reciprocal signals are consistent with the time-reversal-odd symmetry of the MSHE.

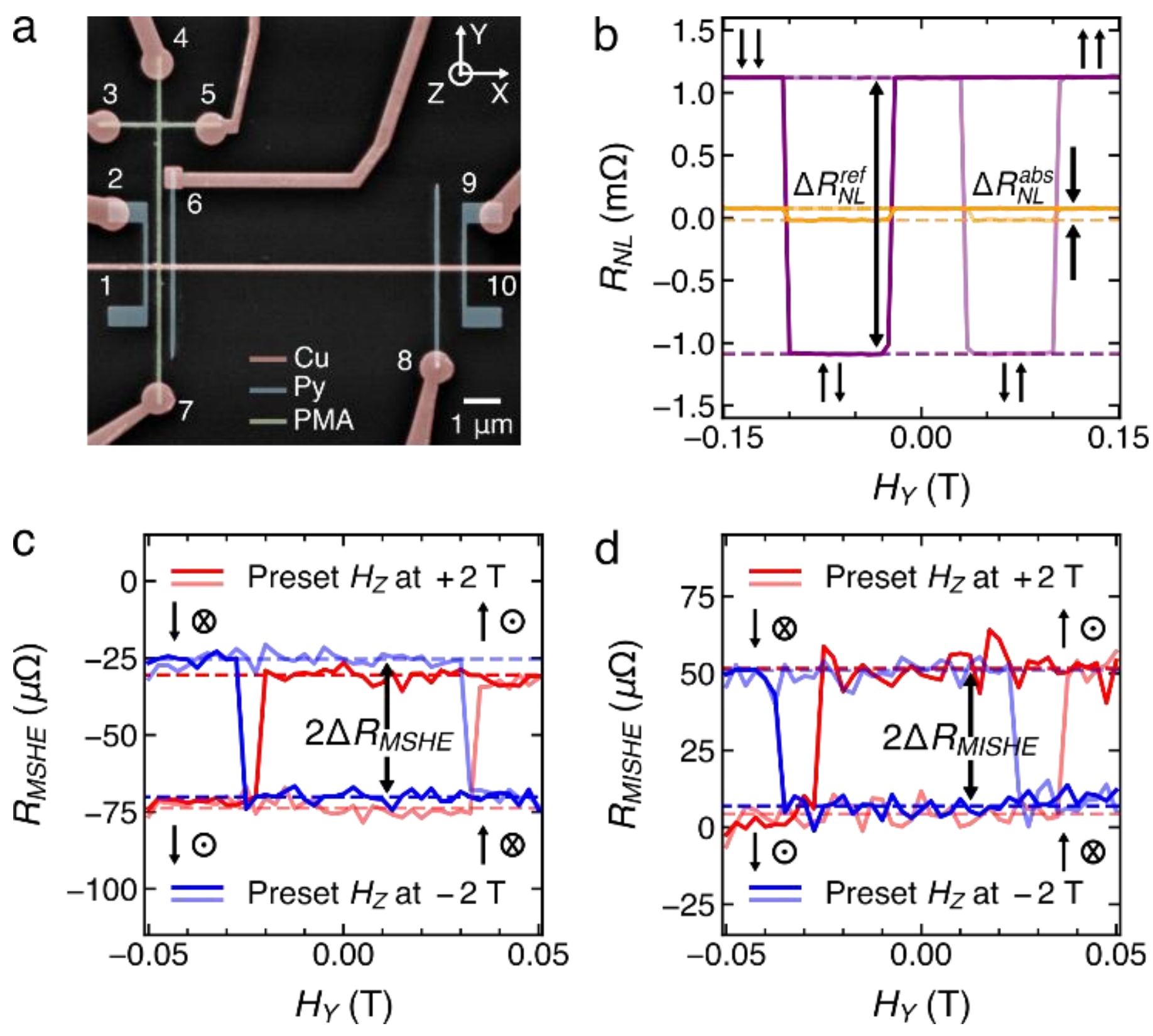

**FIG. 3. Non-local spin signal, MSHE, and MISHE measurements for the lateral spin-valve device at 10 K**. (a) False-colored scanning electron microscope image of the LSV device with the intermediate PMA electrode (left) and the reference LSV device (right). (b) Non-local resistance (detected between contacts 8 and 1) for the reference LSV (purple line) in the biasing geometry across the Py/Cu interface (contacts 9 to 10). The magnetic field is applied along the *y*-direction. The spin signal is denoted as $\Delta R_{NL}^{ref}$. Non-local resistance (detected between contacts 6 and 10) for the LSV with the intermediate PMA electrode (orange line) in the biasing geometry across the Py/Cu interface (contacts 2 to 1). The spin signal is denoted as $\Delta R_{NL}^{abs}$. The vertical arrows indicate the parallel or antiparallel configuration of the Py electrodes. (c) MSHE resistance (detected between contacts 2 and 1) in the biasing geometry along the PMA wire (contacts 4 to 7). Red (blue) curves correspond to measurements after presetting the PMA magnetization with +2 T (-2 T). The dark (light) colors represent the sweep of the magnetic field from +0.05 T (-0.05 T) to -0.05 T (+0.05 T). The arrows indicate the magnetization state of the Py electrode and the PMA electrode. (d) MISHE resistance (detected between contacts 4 and 7) in the biasing geometry across the Py/Cu interface (contacts 2 to 1). Red (blue) curves correspond to measurements after presetting the PMA magnetization with +2 T (-2 T). The arrows in panels c and d indicate the magnetization state of the Py detector electrode and the PMA injector electrode, with the definition of the amplitudes $\Delta R_{MSHE}$ and $\Delta R_{MISHE}$ labeled.

The MSHE is quantified by the magnetic spin Hall conductivity, $\sigma_{MSH}$, and the corresponding magnetic spin Hall angle, $\theta_{MSH} = \sigma_{MSH}/\sigma_{PMA}$, where $\sigma_{PMA}$ is the electrical conductivity of the PMA multilayer. These quantities are determined from the measured $\Delta R_{MSHE}$ and the independently extracted $\lambda_{PMA}^{\perp}$ using Eq. (S2) of the Supp. Info [16]. Measurements from four devices yield an average magnetic spin Hall angle of $\overline{|\theta_{MSH}|} =$ (3.8 ± 0.6)%. Remarkably, this value is of the same order of magnitude as the spin Hall angle $\theta_{SH}$ reported for Pt, a prototypical spin Hall material [17].

Having established the reciprocal behavior of the MSHE and MISHE, we next examine the conventional SHE and ISHE of the PMA multilayer, as shown in Fig. 4. The SHE and ISHE electrical configurations are the same as those used for the MSHE and MISHE, respectively, except that the magnetic field is now swept along the in-plane hard axis of the Py electrode (*x*-axis) to satisfy the symmetry conditions of the conventional SHE and ISHE, rather than those of the MSHE and MISHE (compare Figs. 1a and 1b). The corresponding non-local resistances are thus defined as $R_{(I)SHE} = V_{(I)SHE}/I_{bias}$. Before each field sweep, the magnetization of the PMA multilayer is preset by applying an out-of-plane magnetic field of either $H_Z =$ +2 T (red curve) or -2 T (blue curve), following the same procedure used in the MSHE and MISHE measurements. In contrast to the magnetic spin Hall effects, reversing the PMA magnetization does not reverse the measured ISHE signal, as illustrated by the red and blue curves in Fig. 4a. This magnetization-independent behavior is consistent with the TRS-even character of the conventional SHE (Fig. 1a).

A direct comparison between the ISHE and SHE measurements is presented in Fig. 4b. The reciprocal SHE measurement is performed by interchanging the voltage and current probes relative to the ISHE configuration. The signal amplitudes $\Delta R_{ISHE}$ and $\Delta R_{SHE}$ are defined as the differences between the corresponding resistance values measured for the +*x* and -*x* magnetization states of the Py electrode. In agreement with Onsager-Büttiker reciprocity, the $R_{SHE}$ and $R_{ISHE}$ signals have identical amplitudes but opposite signs. This TRS-even behavior contrasts with that observed for the MSHE and MISHE in Figs. 3c and 3d and further supports the assignment of the latter signals to a TRS-odd contribution to the spin conductivity tensor. Analogously to the MSHE analysis, the spin Hall angle $\theta_{SH}$ is extracted using Eq. (S2) of the Supp. Info [16]. For this quantification we first obtain the relevant spin diffusion length, $\lambda_{PMA}^{\parallel}$, because the applied magnetic field saturates both the Py and PMA magnetizations along *x*, such that the spins absorbed by the PMA electrode are polarized parallel to its magnetization, as discussed in Supp. Info. Note 4. We obtain a value of $|\theta_{SH}| =$ (16.5 ± 0.9)%. Overall, comparison of the extracted spin Hall angles shows that the TRS-odd tensor component responsible for the MSHE reaches (23 ± 4)% of the magnitude of the distinct TRS-even component responsible for the conventional SHE, establishing the MSHE as a truly relevant spin-conductivity response.

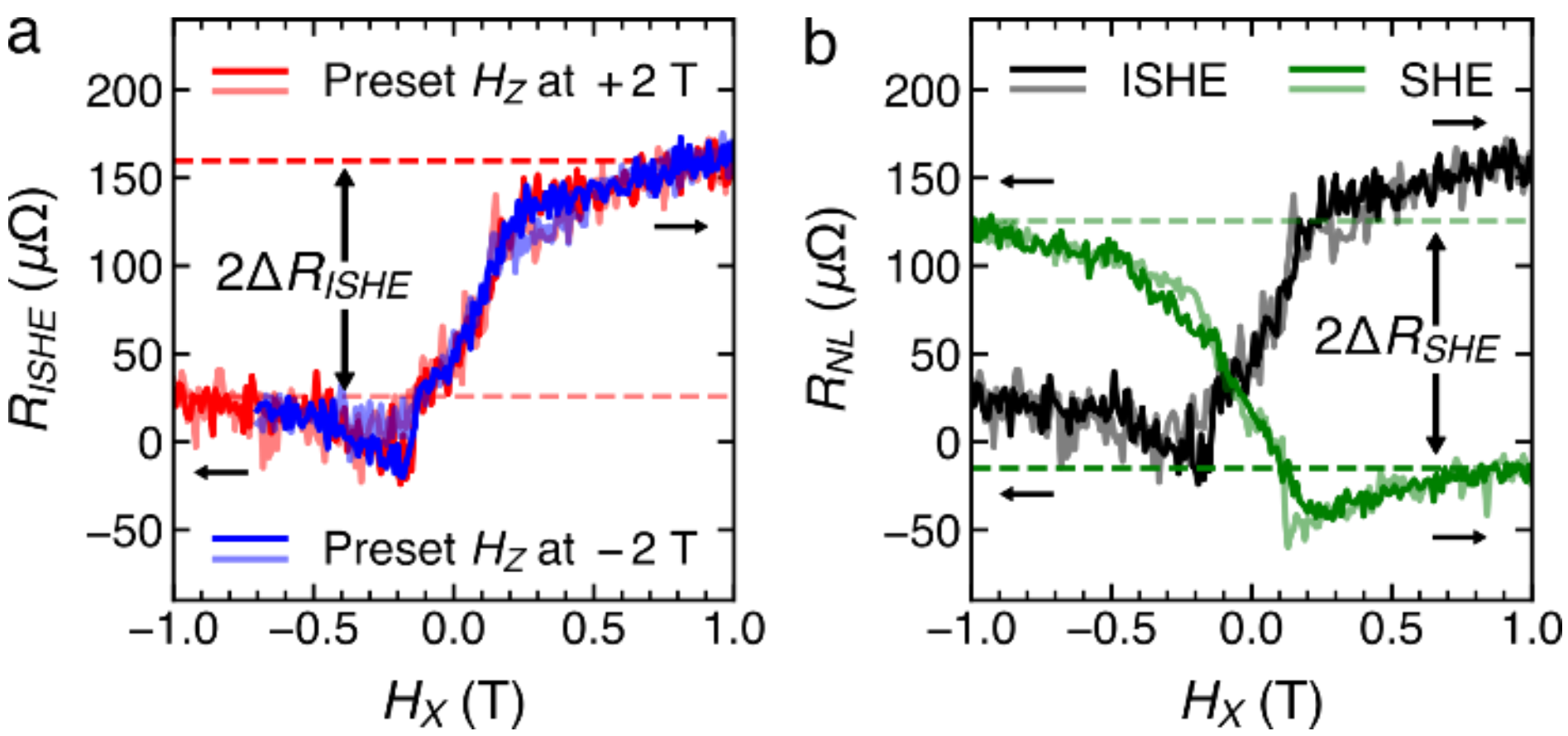

**FIG. 4. SHE and ISHE measurements for the lateral spin-valve at 10 K.** (a) ISHE resistance (detected between contacts 4 and 7, defined in Fig. 3a) in the biasing geometry across the Py/Cu interface (contacts 1 to 2). Red (blue) curves correspond to measurements after presetting the PMA magnetization using a magnetic field of +2 T (-2 T). (b) Comparison between the SHE resistance (green lines), detected between contacts 4 and 7 in the biasing geometry across the Py/Cu interface (contacts 1 to 2), and the ISHE resistance (black lines). The dark (light) colors represent the sweep of the magnetic field from +1 T (-1 T) to -1 T (+1 T). The arrows in both panels indicate the magnetization state of the Py injector (ISHE) and the state of the Py detector (SHE), with the definition of the amplitudes $\Delta R_{ISHE}$ and $\Delta R_{SHE}$ labeled. The magnetic field is applied along the *x*-direction, as defined in Fig. 3a.

In summary, using LSV devices, we have unambiguously identified the magnetic spin Hall effect and its inverse in a perpendicularly magnetized Co-based multilayer. While the MSHE has previously been observed in antiferromagnets [9], our results demonstrate that a sizeable MSHE also arises in a conventional ferromagnet. Reversal of the PMA magnetization reverses the MSHE and MISHE signals, revealing their TRS-odd character, whereas the conventional SHE and ISHE remain unchanged under magnetization reversal, consistent with their time-reversal-even symmetry. These contrasting behaviors enable the electrical isolation of the corresponding TRS-odd and TRS-even components of the spin-conductivity tensor. The MSHE allows configurations beyond the mutually orthogonal relationship among charge current, spin current, and spin polarization imposed by the conventional SHE, thereby expanding the range of spin-current geometries available for spintronic devices. The extracted magnetic spin Hall angle $|\theta_{MSH}| = (3.8 \pm 0.6)\%$ is comparable in magnitude to the conventional spin Hall angle measured in the same PMA multilayer, as well as in heavy metals commonly used in spintronic devices, establishing the MSHE as a large and magnetically switchable contribution to spin-charge interconversion in ferromagnets.

## ACKNOWLEDGMENTS

The authors thank Fernando de Juan for fruitful discussions. We acknowledge funding from MICIU/AEI/10.13039/501100011033 (Grants No. CEX2020-001038-M and CEX2025-001634-M) and from MICIU/AEI and ERDF/EU (Projects No. PID2024-155708OB-I00 and PID2024-155776NB-100). N.D.-G. and J.M. acknowledge funding by the Department of Education of the Basque Government under the Predoctoral Programme for Training of Non-doctoral Research Staff (Fellowships PRE_2024_1_0287 and PRE_2022_1_0297, respectively).

# Supplementary Information

## Note 1. Growth process of multilayers with perpendicular magnetic anisotropy

Thin films were deposited using a magnetron sputtering system (AJA International) at room temperature under a pure atmosphere with a base pressure corresponding to 3 mTorr. The deposition rates were calibrated prior to growth: Pt was deposited at 40 W with a rate of 0.0272 nm·s$^{-1}$; Co, at 50 W with a rate of 0.0305 nm·s$^{-1}$; and Cr, at 40 W with a rate of 0.0328 nm·s$^{-1}$. The $SiO_2$ deposition was carried out at 200 W, although its deposition rate was not determined. The multilayer stack was grown on a Si/$SiO_2$ substrate (n+-doped Si, with a 300-nm-thick thermally grown oxide layer), starting with a 3 nm Cr seed layer, followed by a 3 nm Pt layer. Subsequently, a [Co (0.65 nm)/Pt (0.8 nm)] bilayer was deposited and repeated 10 times to form the multilayer structure with perpendicular magnetic anisotropy (PMA). Finally, a 2 nm Pt capping layer and approximately 10 nm of $SiO_2$ were deposited to prevent the oxidation and contamination of the sample.

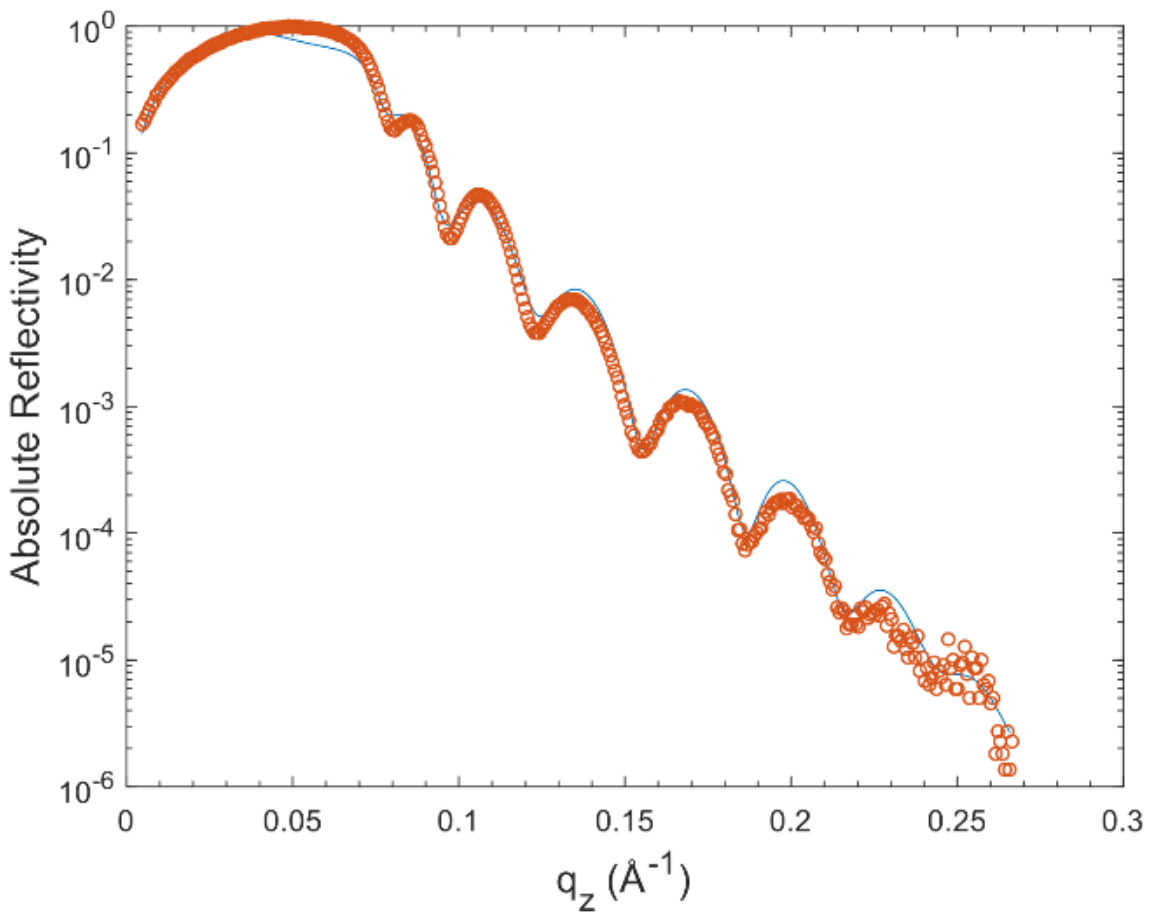

**Fig. S1. X-ray reflectometry spectrum of the PMA thin film before lithography.** The data has been fitted using the Reflex [1] software, confirming the thickness of the different layers present in the sputtered thin film.

## Note 2. Spin transport equations and extraction of the spin transport and magnetic spin Hall parameters

The procedure to extract $\lambda_{PMA}$ with the spin absorption technique is based on comparing the spin signal obtained from the LSV that contains the middle PMA wire with the spin signal obtained from the reference LSV without the PMA. Both devices need to have equivalent Py and Cu electrodes and the same distance, $L$, between Py electrodes. The ratio between both spin signals, $\eta = \Delta R_{NL}^{abs}/\Delta R_{NL}^{ref}$, is given by [2]:

$$\eta = \frac{2Q_{PMA}\left[2Q_{Py}e^{\frac{L}{\lambda_{Cu}}} + 2Q_{Py}^2 e^{\frac{L}{\lambda_{Cu}}} + \sinh\left(\frac{L}{\lambda_{Cu}}\right)\right]}{\cosh\left(\frac{L}{\lambda_{Cu}}\right) - \cosh\left(\frac{L-2d}{\lambda_{Cu}}\right) + 2Q_{Py}\sinh\left(\frac{d}{\lambda_{Cu}}\right)e^{\frac{L-d}{\lambda_{Cu}}} + 2Q_{PMA}\sinh\left(\frac{L}{\lambda_{Cu}}\right) + 4Q_{Py}Q_{PMA}e^{\frac{L}{\lambda_{Cu}}} + 2Q_{Py}\sinh\left(\frac{L-d}{\lambda_{Cu}}\right)e^{\frac{d}{\lambda_{Cu}}} + 2Q_{Py}^2 e^{\frac{L}{\lambda_{Cu}}} + 4Q_{Py}^2 Q_{PMA}e^{\frac{L}{\lambda_{Cu}}}} \quad \text{(S1)}$$

where $Q_{Py(PMA)} = R_{Py(PMA)}/R_{Cu}$, being $R_{Cu} = \lambda_{Cu}\rho_{Cu}/w_{Cu}t_{Cu}$, $R_{Py} = \lambda_{Py}\rho_{Py}/w_{Cu}w_{Py}(1-\alpha_{Py}^2)$, and $R_{PMA} = \lambda_{PMA}\rho_{PMA}/w_{Cu}w_{PMA}\tanh(t_{PMA}\lambda_{PMA})$ the spin resistances of the Cu channel, Py electrodes, and PMA wire, respectively. $\rho_{Cu,Py,PMA}$, $\lambda_{Cu,Py,PMA}$, $w_{Cu,Py,PMA}$, and $t_{Cu,PMA}$ are the resistivities, spin diffusion lengths, widths, and thicknesses, respectively. $\alpha_{Py}$ is the spin polarization of Py. $L$ is the edge-to-edge distance between the two Py electrodes, while $d$ is the edge-to-edge distance between the Py injector and the PMA wire. $\lambda_{Py}$, $\lambda_{Cu}$ and $\alpha_{Py}$ values are well known from previous work [2,3].

Analogous to the ISHE [4], the relation between the experimentally measured $\Delta R_{MISHE}$ and the magnetic spin Hall conductivity ($\sigma_{MSH}$) and magnetic spin Hall angle ($\theta_{MSH}$) is given by:

$$\sigma_{MSH} = \sigma_{PMA}^2 \frac{w_{PMA}}{x_{PMA}} \left(\frac{I_C}{\overline{I_S}}\right) \Delta R_{MISHE} = \sigma_{PMA}\theta_{MSH} \tag{S2}$$

where $x_{PMA}$ is the shunting factor that accounts for the fraction of the current in the PMA wire that is shunted through the Cu. It is obtained from finite-element method (FEM) simulations performed using COMSOL Multiphysics [5]. These FEM simulations use a 3D model geometry that reproduces the Cu and PMA electrodes at their interface area including the dimensions and resistivities. The Cu/PMA interface resistance is considered transparent, meaning no contact impedance is included in the model. A current of 1 μA is applied to the PMA electrode and a finer mesh is used to resolve the local current density. Because the MISHE signal is a voltage measurement along the PMA electrode that depends on the generated current in the PMA electrode, the *y*-component of the current density ($j_y$) is of particular importance. Therefore, the volume integral of $j_y$ is of particular importance. Therefore, the volume integral of $j_y$ is extracted for the Cu electrode and PMA electrode separately and the corresponding current is calculated by: $I_i = \frac{1}{w_{Cu}} \iiint j_y^i dV_i$ with $i$ = Cu, PMA. The electrical shunting is then obtained from $x_{PMA} = I_{PMA}/(I_{Cu} + I_{PMA})$. $\overline{I_S}$ is the effective spin current that contributes to the MISHE in PMA FM and is given by [6]:

$$\frac{\overline{I_S}}{I_C} = \frac{\lambda_{PMA}\left(1 - e^{-\frac{t_{PMA}}{\lambda_{PMA}}}\right)^2}{t_{PMA}\left(1 - e^{-\frac{2t_{PMA}}{\lambda_{PMA}}}\right)} \times \frac{2\alpha_{Py}\left[Q_{Py}\sinh\left(\frac{L-d}{\lambda_{Cu}}\right) + Q_{Py}^2 e^{\frac{L-d}{\lambda_{Cu}}}\right]}{\cosh\left(\frac{L}{\lambda_{Cu}}\right) - \cosh\left(\frac{L-2d}{\lambda_{Cu}}\right) + 2Q_{Py}\sinh\left(\frac{d}{\lambda_{Cu}}\right)e^{\frac{L-d}{\lambda_{Cu}}} + 2Q_{PMA}\sinh\left(\frac{L}{\lambda_{Cu}}\right) + 4Q_{Py}Q_{PMA}e^{\frac{L}{\lambda_{Cu}}} + 2Q_{Py}\sinh\left(\frac{L-d}{\lambda_{Cu}}\right)e^{\frac{d}{\lambda_{Cu}}} + 2Q_{Py}^2 e^{\frac{L}{\lambda_{Cu}}} + 4Q_{Py}^2 Q_{PMA} e^{\frac{L}{\lambda_{Cu}}}} \tag{S3}$$

Using the equations above, we have extracted the spin transport and magnetic spin Hall parameters of the fabricated devices, and the results are summarized in Table S1. The widths of the electrodes and edge-to-edge distances between them have been extracted from SEM images, obtaining $w_{Cu}$ = 124 nm, $w_{PMA}$ = 99 nm, $w_{Py}$ = 118 nm, $L$ = 582 nm, $d$ = 207 nm. The nominal thicknesses are $t_{Cu}$ = 60 nm, $t_{PMA}$ = 20.5 nm and $t_{Py}$ = 30 nm. The resistivities of both Cu and the PMA have been measured as a function of temperature as shown in Fig. S1. The values at 10 K are $\rho_{Cu}$ = 3.192 μΩ·cm and $\rho_{PMA}$ = 70.488 μΩ·cm. Using these values and the measured $\eta$ and $R_{MSHE}$, we extract $\lambda_{PMA}$ (using Eq. (S1)), $|\theta_{MSH}|$ and $|\sigma_{MSH}|$ (using Eqs. (S2 and S3)) for all the devices (D1-D4), summarized in Table S1.

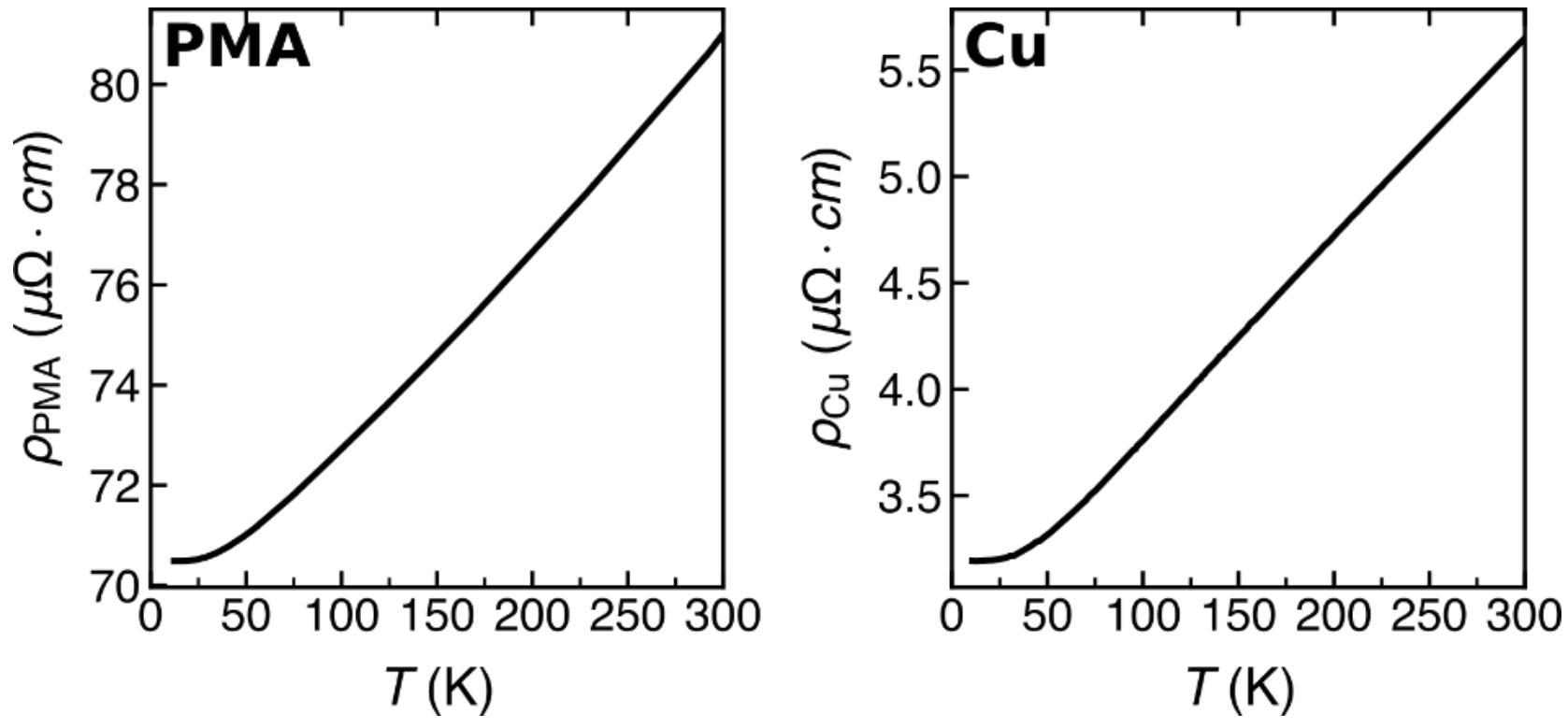

**Fig. S2. Resistivity of the PMA (left) and Cu (right) wires as a function of temperature.**

| Device | $\lambda_{PMA}^{\perp}$ (nm) | $\|\theta_{MSH}\|$ (%) | $\|\sigma_{MSH}\|$ $(\Omega^{-1} \cdot cm^{-1})$ |
|---|---|---|---|
| D1 | 0.473 ± 0.002 | 3.3 ± 0.1 | 462 ± 11 |
| D2 | 0.293 ± 0.002 | 4.45 ± 0.13 | 566 ± 15 |
| D3 | 0.230 ± 0.002 | 4.90 ± 0.14 | 695 ± 14 |
| D4 | 0.240 ± 0.003 | 2.35 ± 0.10 | 321 ± 14 |
| Average | 0.31 ± 0.06 | 3.8 ± 0.6 | 510 ± 80 |

**Table S1. Spin transport and magnetic spin Hall parameters of the fabricated devices.** Calculated values of the spin diffusion length ($\lambda_{PMA}^{\perp}$), magnetic spin Hall angle ($\theta_{MSH}$) and magnetic spin Hall conductivity ($\sigma_{MSH}$) at 10 K for each device. The average value of each parameter is presented in the last row. The errors associated with the individual values are calculated by propagating the errors of the parameters included in Eqs. (S1-S3). The errors associated with the average values correspond to the standard error of the mean.

### Note 3. Reciprocity in the magnetic spin Hall effect

The sign of the non-local resistance upon exchanging current and voltage terminals in a multiterminal device can be understood from Onsager-Büttiker reciprocity together with the symmetry properties of the spin Hall conductivity tensor $\sigma_{ij}^{S_k}$. In linear response, the reciprocity relation reads

$$R_{ab;cd}\,(B, \boldsymbol{M}) = R_{cd;ab}\,(-B, -\boldsymbol{M}),$$

where $a$, $b$ denote current terminals; $c$, $d$ voltage terminals; and $\boldsymbol{M}$ is the magnetization of the ferromagnetic element. This relation follows from Onsager symmetry extended to multiterminal conductors [7,8].

The non-local signal measured in spin-charge conversion experiments can be generically written as

$$R_{nl} \propto P_{\alpha}\, D\sigma_{ij}^{S_k},$$

where $P_{\alpha}$ is the spin polarization of the ferromagnetic injector/detector, $D$ accounts for spin diffusion, and $\sigma_{ij}^{S_k}$ is the spin Hall tensor relating a charge current $j_j$ to a spin current $J_i^{S_k}$.

The crucial point is that different tensor components have different transformation properties under time reversal, which determines the observed sign behavior:

<u>(i) Conventional SHE: $i \neq j \neq k$</u>

For the conventional spin Hall effect, the tensor component $\sigma_{ij}^{S_k}$ is of Hall type, with all indices different. These components are even under time reversal, i.e.

$$\sigma_{ij}^{S_k}\,(\boldsymbol{M}) = \sigma_{ij}^{S_k}\,(-\boldsymbol{M}).$$

Applying reciprocity to the non-local measurement,

$$R_{ab;cd}^{SHE}\,(B, \boldsymbol{M}) = R_{cd;ab}^{SHE}\,(-B, -\boldsymbol{M}) = R_{cd;ab}^{SHE}\,(-B, \boldsymbol{M}),$$

since the tensor does not depend on the sign of $\boldsymbol{M}$. Upon mapping back to the original field configuration (i.e., comparing signals at the same B), this results in an overall sign reversal when exchanging current and voltage terminals, reflecting the interchange between direct (SHE) and inverse (ISHE) processes.

At the same time, because the tensor is TRS-even, the signal does not change sign under magnetization reversal.

(ii) Magnetic SHE: $k = i$ or $k = j$

In a ferromagnetic system, additional tensor components are allowed, such as

$$\sigma_{ij}^{s_j} \text{ or } \sigma_{ij}^{s_i},$$

which correspond to the magnetic spin Hall effect. These components are odd under time-reversal, since they are proportional to the magnetization:

$$\sigma_{ij}^{k}(\boldsymbol{M}) = -\sigma_{ij}^{k}(-\boldsymbol{M}).$$

Applying reciprocity now yields

$$R_{ab;cd}^{MSHE}(B, \boldsymbol{M}) = R_{cd;ab}^{MSHE}(-B, -\boldsymbol{M}).$$

In contrast to the conventional case, reversing $\boldsymbol{M}$ flips the sign of the tensor, compensating for the sign change associated with probe exchange. As a consequence,

$$R_{ab;cd}^{MSHE}(B, \boldsymbol{M}) = R_{cd;ab}^{MSHE}(B, \boldsymbol{M}),$$

and the non-local resistance retains its sign when current and voltage terminals are swapped. However, due to the explicit dependence on $\boldsymbol{M}$, the signal changes sign upon magnetization reversal,

$$R_{ab;cd}^{MSHE}(B, \boldsymbol{M}) = -R_{ab;cd}^{MSHE}(B, -\boldsymbol{M}),$$

in agreement with the odd time-reversal character of these tensor components.

**Note 4. Anisotropy of the spin diffusion length**

To investigate the anisotropy of spin absorption in the PMA ferromagnet, the non-local spin signal was measured with the magnetic field applied along both the easy axis (*y*-direction) and the hard axis (*x*-direction) of the electrodes. In a ferromagnetic material, the spin diffusion length can depend on the relative orientation between the spin polarization of the spin current and the magnetization of the material [9]. In the measurement shown in Fig. 3b of the main text, a small magnetic field is applied along the easy axis of the Py electrodes. This field switches the Py magnetizations without rotating the PMA magnetization away from the out-of-plane direction. Consequently, the injected spins are polarized perpendicular to the PMA magnetization, and the measurement probes the transverse spin diffusion length, $\lambda_{PMA}^{\perp}$. In contrast, a sufficiently large magnetic field applied along the hard axis of the Py electrode aligns both the Py and PMA magnetizations along the *x*-direction. The injected spins are then polarized parallel to the PMA magnetization, allowing the longitudinal spin diffusion length, $\lambda_{PMA}^{\parallel}$, to be determined. The extraction of $\lambda_{\mathrm{PMA}}^{\parallel}$ requires a comparison between the easy-axis and hard-axis measurements, as described below.

Fig. S3 shows the measurements performed on device D5, the same device used for the SHE and ISHE measurements presented in the main text. Figs. S3a and S3b show the non-local resistance of the reference

LSV and that of the LSV containing the PMA absorber, respectively. Each panel includes measurements with the magnetic field applied along both the easy and hard axes of the Py electrodes.

For the perpendicular configuration, the spin-signal amplitudes are obtained directly from the difference between the parallel and antiparallel resistance states measured along the easy axis, following the same procedure described for Fig. 3b of the main text. Introducing the ratio of the amplitudes into Eq. (S1) yields $\lambda_{\mathrm{PMA}}^{\perp} = 0.203 \pm 0.005$ nm for device D5.

For the parallel configuration, the spin-signal amplitudes are extracted by combining the easy-axis and hard-axis measurements. Because the measurements are performed on the same LSV, they share a common resistance baseline. Therefore, the average easy-axis resistance corresponds to the midpoint between the parallel and antiparallel resistance states, and the spin-signal amplitude is obtained as twice the difference between this average value and the saturated resistance measured along the hard axis. Using the resulting ratio in Eq. (S1), we obtain $\lambda_{\mathrm{PMA}}^{\parallel} = 0.283 \pm 0.013$ nm. This value is used to extract the conventional spin Hall angle reported in Table S2. The difference between $\lambda_{\mathrm{PMA}}^{\parallel}$ and $\lambda_{\mathrm{PMA}}^{\perp}$ within the same device demonstrates anisotropic spin absorption in the PMA multilayer, consistent with previous reports for other ferromagnets [3].

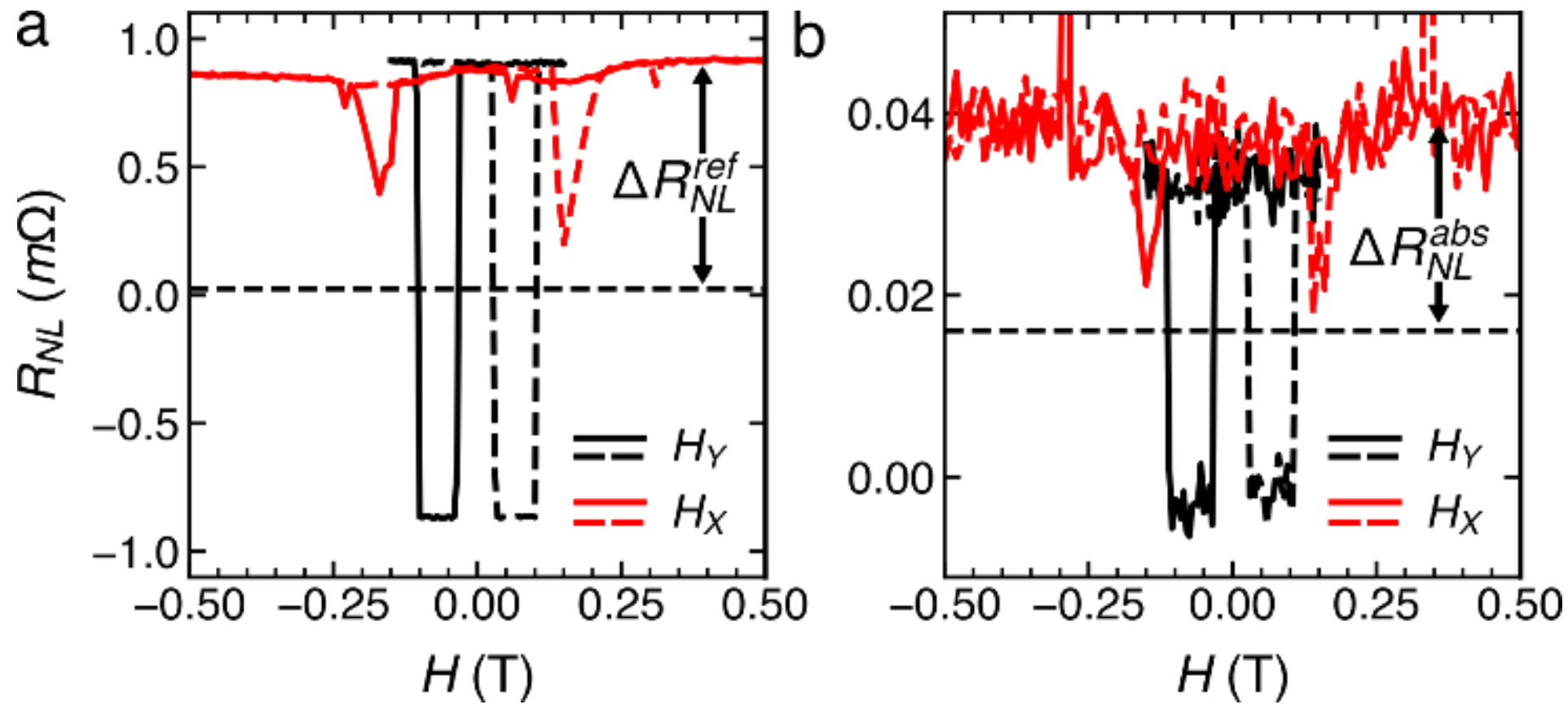

**Fig. S3. Magnetic-field-direction dependence of the spin signal for device D5.** (a) Non-local resistance for the reference Py/Cu/Py LSV sweeping the magnetic field in the hard axis (red curve) and in the easy axis of the Py (black curve). (b) Non-local resistance for the absorption Py/Cu/PMA LSV sweeping the magnetic field in the hard axis (red curve) and in the easy axis of the Py (black curve).

| Device | $\lambda_{PMA}^{\parallel}$ (nm) | $\lvert\theta_{SH}\rvert$ (%) | $\lvert\sigma_{SH}\rvert$ $(\Omega^{-1} \cdot cm^{-1})$ |
|---|---|---|---|
| D5 | 0.283 ± 0.013 | 16.5 ± 0.9 | 2250 ± 120 |

**Table S2. Spin transport and spin Hall parameters of device 5.** Calculated values of the spin diffusion length ($\lambda_{PMA}^{\parallel}$), spin Hall angle ($\theta_{SH}$) and spin Hall conductivity ($\sigma_{SH}$) at 10 K. The errors associated with the individual values are calculated by propagating the errors of the parameters included in Eqs. (S1-S3).

## SUPPLEMENTARY REFERENCES